\documentclass[12pt,letter]{article}

\usepackage{amsmath,amsthm,latexsym,amssymb,amsfonts,amsbsy,mathrsfs,bm,braket,color,psfrag,epsf}
\usepackage[mathscr]{eucal}
\usepackage{graphicx,multirow,tabularx}
\usepackage{amsfonts,delarray}

\usepackage[comma]{natbib}
\usepackage{url}

\usepackage{algorithm}
\usepackage{algorithmic}

\usepackage{fancyhdr}
\usepackage[tight,footnotesize]{subfigure}
\makeatletter
\renewcommand\section{\@startsection {section}{1}{\z@}%
                                   {-3.5ex \@plus -1ex \@minus -.2ex}%
                                   {2.3ex \@plus.2ex}%
                                   {\centering\normalfont\Large\bf\MakeUppercase}}
\makeatother

\newcommand{\blind}{0}

\numberwithin{equation}{section}
\newcommand{\bSigma}{\boldsymbol{\Sigma}}
\newcommand{\bpsi}{\boldsymbol{\psi}}
\newcommand{\bPsi}{\boldsymbol{\Psi}}
\newcommand{\bLambda}{\boldsymbol{\Lambda}}
\newcommand{\bt}{\bm t}

\newcommand{\bb}{\bm b}
\newcommand{\bQ}{\bm Q}
\newcommand{\bI}{\bm I}
\DeclareMathOperator*{\argmin}{arg\,min}

\newtheorem{thm}{Theorem}[section]
\newtheorem{cor}{Corollary}[section]

\begin{document}


\def\spacingset#1{\renewcommand{\baselinestretch}%
{#1}\small\normalsize} \spacingset{1}

\if0\blind
{
  \title{\bf Multi-dimensional Functional Principal Component Analysis}
  \author{Lu-Hung Chen\footnote{Lu-Hung Chen is Assistant Professor, Institute of Statistics, National Chung-Hsing University, Taichung 402, Taiwan. Email: luhung@nchu.edu.tw.}
    and 
    Ci-Ren Jiang\footnote{Ci-Ren Jiang is Assistant Research Fellow, Institute of Statistical Science, Academia Sinica, Taipei 115, Taiwan. Email: cirenjiang@stat.sinica.edu.tw.}} 
    
  \maketitle
} \fi

\if1\blind
{
  \bigskip
  \bigskip
  \bigskip
  \begin{center}
    {\LARGE\bf Multi-dimensional Functional Principal Component Analysis}
\end{center}
  \medskip
} \fi

\bigskip

\begin{abstract}
Functional principal component analysis is one of the most commonly employed approaches in functional and longitudinal data analysis and we extend it to analyze functional/longitudinal data observed on a general $d$-dimensional domain. The computational issues emerging in the extension  are fully addressed with our proposed solutions. The local linear smoothing technique is employed to perform estimation because of its capabilities of performing large-scale smoothing and of handling data with different sampling schemes (possibly on irregular domain) in addition to its nice theoretical properties.  Besides taking the fast Fourier transform strategy in smoothing, the modern GPGPU (general-purpose computing on graphics processing units) architecture is applied to perform parallel computation to save computation time. To resolve the out-of-memory issue due to large-scale data, the random projection procedure is applied in the eigendecomposition step. We show that the proposed estimators can achieve the classical nonparametric rates for longitudinal data and the optimal convergence rates for functional data if the number of observations per sample is of the order $(n/ \log n)^{d/4}$. Finally, the performance of our approach is demonstrated with simulation studies and the fine particulate matter (PM 2.5) data measured in Taiwan. 
\end{abstract}

\noindent%
\vfill

\newpage

\section{Introduction}\label{intro}

Functional principal component analysis (FPCA), inherited from principal component analysis (PCA), has been widely employed in various fields of functional data analysis (FDA) and longitudinal data analysis; e.g., functional regression \citep{YaoMW:05b,GertGCG:13}, functional classification \citep{LengM:06,Chiou:12}, functional clustering \citep{ChiouL:07,ChiouL:08}, outlier detection \citep{Gervini:09}, survival analysis \citep{ChioM:09}, time series analysis \citep{HyndmanS:09,AueDH:14}, etc. Its applications also cover a wide range of topics (e.g., \cite{LengM:06,ChioM:09,LiuM:09,AstoCE:10,Chiou:12,JianAW:15}, etc.), and various estimation methods have been proposed for FPCA: \cite{RiceS:91} and \cite{Silverman:96} represented the latent functions with B-splines and extracted the eigenfunctions by performing generalized eigendecomposition owing to the roughness penalty; \cite{JameHS:00} and \cite{RiceW:01} adopted the mixed effect models where the mean function and the eigenfunctions were represented with B-splines and the spline coefficients were estimated by the EM algorithm; \cite{YaoMW:05a} applied the local linear smoothers \citep{FanG:96} to estimate the mean and the covariance functions and obtained the eigenfunctions by solving the corresponding eigen-equations. The principal component scores were predicted by the conditional expectation approach, named ``PACE''. Some computational tricks for densely observed functional data can be found in \citep{ChenZPM:15} to accelerate the computation.  
 
Even though FDA has received considerable attention over the last decade, 
most approaches still focus on one-dimensional functional data. Few are developed for general $d$-dimensional functional data. To extend FPCA to its multi-dimensional version, in practice one has to take good care of the following issues. \textbf{Issue A} is that the conventional strategy for obtaining eigenfunctions needs to be modified as the covariance of $d$-dimensional functional data becomes a $2d$-dimensional function instead of a matrix. 
\textbf{Issue B} is the enormous size of an empirical covariance function and applying any computation to an object of this scale is demanding. 
The last issue, \textbf{Issue C}, is that the $d$-dimensional domain may not be rectangle which could cause challenging problems in many smoothing approaches. 

Recently, some attention has been drawn to spatial and image data in the Statistics society (e.g., \cite{ZhuZIP:07,AstoK:12,TianHSL:12,ZhanLBC:13,RiskMREC:14}, etc.). In particular, several attempts have been made to extend FPCA for spatial and image data. \cite{ZipunnikovCYDSC11a,ZipunnikovCYDSC11b} vectorized the pre-smoothed images and extracted the eigenfunctions by singular value decomposition (SVD). \cite{WangH15} proposed a regularized SVD approach for two-dimensional spatial data. However, SVD could only be applied to functional data observed on a regular grid. To handle bivariate functional observations made on irregular domain, \cite{Zhoup14} extended the functional mixed effect models \citep{JameHS:00,RiceW:01} to bivariate functional data by utilizing the triangularized bivariate splines \citep{ChuiL:87,LaiW:13}. However, the triangularized bivariate splines are designed for two-dimensional functions only. Extending spline basis functions for general $d$-dimensional data observed on an irregular domain is very sophisticated and becomes extremely complex as $d$ increases.  \cite{HungWTH12} applied the multilinear PCA (MPCA) \citep{LuPV08} to analyze two-dimensional facial images. MPCA requires less computational resources and is much faster than general $d$-dimensional FPCA, referred as d-FPCA hereafter. However, the eigentensors obtained by MPCA are usually uninterpretable. Moreover, MPCA is not applicable if the $d$-dimensional functions are not observed on rectangle meshes.

We will employ the local linear smoothing technique to perform estimation in d-FPCA in this paper because of its nice theoretical properties \citep{HallH:06,HallMW:06,LiH:10:1} and its capability of handling data with different sampling schemes over various domains. Thus, \textbf{Issue C} is solved. 
A local linear smoother is also more appropriate for large-scale smoothing as most smoothing techniques, such as spline-based approaches, require the users to solve a global linear system where all the data are involved. However, for large-scale problems loading all the data into a computer's memory is nearly impossible. On the contrary, a local linear smoother estimates the function locally and independently, and thus can be performed parallelly and distributively; for example, it can be done through a modern GPGPU (general-purpose computing on graphics processing units) architecture. Another nice aspect of a local linear smoother is that it can easily be incorporated with the random projection procedure \citep{HalkoMT:11} and block matrix operations, which make the eigendecomposition step feasible when the empirical covariance function is of an enormous size. When the observations are made at the nodes of a regular grid, \textbf{Issue C} no longer exists and the local linear smoothing procedure can be further accelerated by using the FFT (fast Fourier transform) based approaches (e.g., \cite{Silverman:82,Breslaw:92,Wand:94}, etc.). These FFT based approaches can be easily parallelized as well and these arguments regarding to computational issues will be further elaborated in Section 3.

The rest of this paper proceeds as follows. Section 2 is comprised of the d-FPCA framework, the proposed estimators and their asymptotic properties. The major computational issues with solutions are provided in Section 3. Simulation studies and an illustrative real data analysis are presented in Sections 4 and 5, respectively. Section 6 contains concluding remarks. The theoretical assumptions are delegated to an appendix.

\section{Methodology}

We consider the stochastic process, for $\bm t\in \Omega$,
\begin{equation}
	X(\bm t)=\mu(\bm t)+\sum_{\ell=1}^{\infty}A_{\ell}\phi_{\ell}(\bm t),
\label{KLdecomposition}
\end{equation}
where $\mu(\bm t)$ is the mean function, $\phi_{\ell}(\bm t)$ is the $\ell$-th eigenfunction of $\Gamma(\bm s,\bm t)$, the covariance function of $X(\bm t)$, and $A_\ell$ is the $\ell$-th principal component score. Without loss of generality, the domain of $X(\bm t)$ is assumed to be a compact space $\Omega\subset [0,1]^d$ and thus the domain of $\Gamma(\bm s, \bm t)$ is $\Omega\times\Omega$. The principal component scores, $A_{\ell}$'s, are uncorrelated random variables with mean 0 and variance $\lambda_{\ell}$, where $\lambda_{\ell}$ is the eigenvalue of $\Gamma(\bm s, \bm t)$ corresponding to eigenfunction $\phi_{\ell}(\bm t)$. We further assume that $\mu(\bm t )$, $\Gamma(\bm s, \bm t)$ and $\phi_{\ell}(\bm t)$ are smooth functions. Detailed assumptions can be found in the Appendix.

\subsection{Estimation}
In practice, most data are available at discrete time points and might further be contaminated with measurement errors. We denote the observations of the $i$-th sample made at $\bm t_{ij}=(t_{ij,1},\ldots,t_{ij,d})^T$ as  
\begin{equation}
\begin{split}
	Y_{ij}&=X_{i}(\bm t_{ij})+\epsilon_{ij} \\
		&=\mu(\bm t_{ij})+\sum_{\ell=1}^{\infty}A_{i\ell}\phi_{\ell}(\bm t_{ij})+\epsilon_{ij},
\end{split}
\label{dataprocess}
\end{equation}
where $1\leq j \leq N_i$, $1\leq i\leq n$, and $\epsilon_{ij}$ is the random noise with mean $0$ and variance $\sigma^2$.

In order to reconstruct $X_i(\bm t)$ in (\ref{dataprocess}) at any given $\bm t = (t_1,\ldots, t_d)^T \in \Omega$, we need to estimate $\mu(\bm t)$, $\Gamma(\bm s, \bm t)$ and $\sigma^2$. When the observations are not made on a regular grid, conventional approaches, where functional data are treated as high dimensional data, can not be directly applied. Statistical approaches with smoothing techniques are often considered; here we employ the local linear smoothers.  Specifically, the mean function is estimated by 
\begin{equation}
\begin{split}
 \hat\mu( \bm t) =  & \hat b_0, \\
 \hat{\bm b} =  & \argmin_{\bm b = (b_0,\ldots,b_d)^T \in R^{d+1}} \sum_{i=1}^n \frac{1}{N_i} \sum_{j=1}^{N_i} K_h(\bm t - \bm t_{ij}) \\ & \times [Y_{ij}-b_0-\sum_{k=1}^d b_k(t_k-t_{ij,k})]^2,
\end{split}
\label{eq:muhat}
\end{equation}
where 
\[
K_{h_\mu}(\bm t - \bm t_{ij}) =  K\left( \frac{t_1-t_{ij,1}}{h_{\mu,1}},\ldots,\frac{t_d-t_{ij,d}}{h_{\mu,d}} \right)\prod_{k=1}^d \frac{1}{h_{\mu,k}},
\]
$h_{\mu,k}$ is the bandwidth for the $k$-th coordinate and $K(\cdot)$ is a $d$-dimensional kernel function. Once the mean function is estimated, the covariance function can be obtained by 
\begin{equation}
\begin{split}
 \hat\Gamma( \bm s, \bm t)  = & \hat b_0 - \hat\mu(\bm s)\hat\mu(\bm t), \text{ where}\\
\hat{\bm b} = & \argmin_{\bm b = (b_0,\ldots,b_{2d})^T \in R^{2d+1}}   \sum_{i=1}^n \frac{1}{N_i(N_i-1)} \\ & \times \sum_{1\leq j\neq \ell \leq N_i} K_{h_\Gamma}(\bm s - \bm s_{ij})K_{h_\Gamma}(\bm t - \bm t_{i\ell}) \\
& \times \left[Y_{ij}Y_{i\ell}-b_0-\sum_{k=1}^d b_k(s_k-s_{ij,k})\right. \\ & \left. -\sum_{k=1}^d b_{d+k}(t_{k}-t_{i\ell,k})\right]^2,
\end{split}
\label{eq:Gammahat}
\end{equation}
$h_{\Gamma,k}$'s are the bandwidths and $K(\cdot)$ is defined as (\ref{eq:muhat}). The eigenvalues and eigenfunctions can be estimated by solving (\ref{eigendecompose}) with $\Gamma(\bm s, \bm t)$ replaced by $\hat\Gamma(\bm s, \bm t)$. Details could be found in Section 3.

When $Y_{ij}$'s are densely observed, the principal component scores, $A_{i\ell}$'s, can be predicted by inner products. However, when the observations are not dense enough, applying inner products might not generate satisfactory results. To solve this issue for irregularly observed longitudinal data, \cite{YaoMW:05a} proposed PACE. It is known that the PACE is a BLUP (best linear unbiased predictor) even when the normality assumption is violated. To employ PACE, we further need the estimate of $\sigma^2$. Since $\text{cov}(Y_{ij},Y_{ij'}) = \Gamma(\bm t_{ij}, \bm t_{tj'}) + \sigma^2 \delta_{jj'}$, where $\delta_{jj'}=1$ if $j=j'$ and 0 otherwise, $\sigma^2$ can be estimated by the strategy taken in \cite{YaoMW:05a}. First, we estimate $\Gamma(\bm t, \bm t)+\sigma^2$ by $\hat b_0$, where
\begin{align}
\hat{\bm b}   =  & \argmin_{\bm b = (b_0,\ldots,b_d)^T \in R^{d+1}}  \sum_{i=1}^n \frac{1}{N_i} \sum_{j=1}^{N_i} K_{h_\sigma}(\bm t - \bm t_{ij}) \nonumber \\ & \times \left[Y_{ij}^2-b_0-\sum_{k=1}^d b_k(t_k-t_{ij,k})\right]^2, 
\label{eq:sigma2hatsmooth}
\end{align}
$K_{h_\sigma}$ is defined as (\ref{eq:muhat}) and $h_{\sigma,k}$'s are the bandwidths. Then, $$\hat\sigma^2(\bm t) = \hat b_0 - \hat\Gamma(\bm t, \bm t)+ \hat\mu(\bm t)\hat\mu(\bm t).$$ If $\sigma^2$ does not change over $\bm t$, we consider to estimate it by  
\begin{equation} 
\hat\sigma^2 = \int_\Omega \hat\sigma^2(\bm t) d\bm t.
\end{equation}
Note that more sophisticated approaches could be considered to produce more robust estimates for $\sigma^2$. With $\hat\sigma^2$, we can now predict $A_{i,\ell}$ by PACE. Specifically,   
\begin{equation}
\hat A_{i,\ell} = \hat\lambda_\ell \hat\Phi_{i\ell}^T \hat\Sigma_{\bm Y_i}^{-1} \bm Y^C_i, 
\label{PACE}
\end{equation}
where 
\begin{align*} & \hat\Phi_{i\ell}  = (\hat\phi_\ell(\bm t_{i1}),\ldots,\hat\phi_\ell(\bm t_{iN_i}))^T, \\ 
& \hat\Sigma_{\bm Y_i}  = \sum_{\ell=1}^L \hat\lambda_\ell \hat\Phi_{i\ell}\hat\Phi_{i\ell}^T + \hat\sigma^2 I_{N_i\times N_i}, \\  & \text{and }\bm Y^C_i = \left(Y_i(\bm t_{i1})-\hat\mu(\bm t_{i1}), \ldots, Y_i(\bm t_{iN_i})-\hat\mu(\bm t_{iN_i})\right)^T.
\end{align*} 

By combining the above estimates, $X_i(\bm t)$ can be reconstructed as $$\hat X_i(\bm t) = \hat\mu(\bm t)+\sum_{\ell=1}^{L}\hat A_{i\ell}\hat\phi_{\ell}(\bm t),$$
where $L$ can be selected by AIC, BIC, FVE (fraction of variation explained) or other model selection criteria; FVE is employed in this paper. The product Epanechnikov kernel function is employed in our numerical studies. The bandwidths in the aforementioned estimators can be decided with some data-driven approach, such as the method proposed in \cite{RuppertSW:94}. We will elaborate our bandwidth selection procedure in section 3.3.

\subsection{Asymptotical properties}
The strategies taken in \cite{LiH:10:1} may be employed to show the asymptotical properties of our estimators and similar asymptotic properties are obtained.  Below we will provide the results while the assumptions are listed in the Appendix. Two sampling schemes on $N_i$ for $d\geq 2$ will be discussed; one corresponds to functional data, while the other corresponds to longitudinal data.  Denote $\gamma_{nk} = \left( n^{-1}\sum_{i=1}^n N_i^{-k} \right)^{-1}$ for $k=1$ and $2$, $\delta_{n1}(h)= [\{1 + 1/(h^d\gamma_{n1})\}\log n/n]^{1/2}$ and $\delta_{n2}(h) = [\{1 + 1/(h ^d\gamma_{n1})+1/(h^{2d}\gamma_{n2})\}\log n/n]^{1/2}$.  We first provide the convergence rate for $\hat\mu(\bm t)$.

\begin{thm} \label{A:thm1}
Assume that A.1-A.4 hold. Then,
\begin{equation} \label{eq:muas}
\sup_{\bm t\in\Omega} |\hat{\mu}(\bm t)-\mu(\bm t)| = O(h_\mu^2+ \delta_{n1}(h_\mu)) \text{ } a.s.
\end{equation}
\end{thm}

On the right hand side of (\ref{eq:muas}), $O(h_\mu^2)$ is a bound for bias and the other term is the uniform bound for $|\hat\mu(\bm t) -E(\hat\mu (\bm t))|$. We next investigate the asymptotical results of Theorem \ref{A:thm1} under two special sampling schemes.

\begin{cor} \label{A:cor1}
Assume that A.1-A.4 hold. \\
(a) If $\max_{1\leq i\leq n} N_i \leq \mathcal{M}$ for some fixed $\mathcal{M}$, then
\begin{equation*}
\sup_{\bm t\in\Omega} |\hat{\mu}(\bm t)-\mu(\bm t)| = O(h_\mu^2 + \{\log n/(nh_\mu^{d})\}^{1/2}\} \text{ } a.s.
\end{equation*}
(b) If $\max_{1\leq i\leq n} N_i \geq \mathcal{M}_n$, where \\ $\mathcal{M}_n^{-1} \approx h_\mu^d \approx (\log n/n)^{d/4}$ is bounded away from zero, then
\begin{equation*}
\sup_{\bm t\in\Omega} |\hat{\mu}(\bm t)-\mu(\bm t)| = O( (\log n/n)^{1/2}\} \text{ } a.s.
\end{equation*}
\end{cor}

Corollary \ref{A:cor1} indicates that the the classical nonparametric rate for estimating a $d$-dimensional function can be achieved for longitudinal data and that the optimal convergence rate can be achieved if $N_i$, the number of observations per sample, is of the order $(n/\log n)^{d/4}$.

The following results are the convergence rates for $\hat\Gamma(\bm s, \bm t)$ and $\hat\sigma^2$.

\begin{thm} \label{A:thm2}
Assume that A.1-A.6 hold. Then,
\begin{align*}
\sup_{\bm s,\bm t\in\Omega} & |\hat{\Gamma}(\bm s, \bm t)-\Gamma(\bm s,\bm t)| \\  & = O\left(h_\mu^2 + \delta_{n1}(h_\mu)+  h_\Gamma^2+\delta_{n1}(h_\Gamma)\right) \text{ } a.s.
\end{align*}
\end{thm}

\begin{thm} \label{A:thm3}
Assume that A.1-A.2 and A.5-A.7 hold. Then,
\begin{align*}
& \sup  |\hat\sigma^2-\sigma^2| \\ & = O\left(h_\Gamma^2+\delta_{n1}(h_\Gamma)+\delta_{n2}^2(h_\Gamma)+h_\sigma^2+\delta_{n1}^2(h_\sigma)\right) \text{ } a.s.
\end{align*}
\end{thm}

\begin{cor} \label{A:cor2}
Assume that A.1-A.6 hold. \\
(a) If $\max_{1\leq i\leq n} N_i \leq \mathcal{M}$ for some fixed $\mathcal{M}$ and $h_\Gamma^{1/2} \lesssim h_\mu \lesssim h_\Gamma $ then
\begin{align*}
\sup_{\bm s,\bm t\in\Omega} & |\hat{\Gamma}(\bm s, \bm t)-\Gamma(\bm s, \bm t)| \\ & = O\left( h_\Gamma^2 + \{\log n/(nh_\Gamma^{2d})\}^{1/2}\right) \text{ } a.s.
\end{align*}
(b) If $\max_{1\leq i\leq n} N_i \geq \mathcal M_n$, where $\mathcal M_n^{-1} \lesssim h_\Gamma^d \lesssim h^d_\mu \lesssim (\log n/n)^{d/4}$ is bounded away from zero, then
\begin{equation*}
\sup_{\bm s,\bm t\in\Omega}  |\hat{\Gamma}(\bm s, \bm t)-\Gamma(\bm s, \bm t)|  = O\left((\log n/n)^{1/2}\right) \text{ } a.s.
\end{equation*}
\end{cor}

Corollary \ref{A:cor2} shows that the classical nonparametric rate for estimating a $2d$-dimensional function can be obtained for longitudinal data and the optimal convergence rates can be achieved if $N_i$ is of the order $(n/ \log n)^{d/4}$ for functional data which is the same as the condition required for $\hat\mu(\bm t)$ to have optimal convergence rate.

Further, we assume the nonzero eigenvalues are distinct as \cite{LiH:10:1} due to the identifiability of eigenfunctions.  

\begin{thm}\label{A:thm4}
Assume that A.1-A.6 hold, for $1\leq j\leq J$,
\begin{align*}
 (i) & |\hat\lambda_j -\lambda_j|  \\ &  = O\left( (\log n/n)^{1/2} + \zeta +\delta_{n1}^2(h_\mu) \right) \text{  a.s.,} \\
 (ii) & \|\hat\phi_j(\bm t) -\phi_j (\bm t)\|  \\ &  = O\left(\zeta+ \delta_{n1}(h_\mu)+\delta_{n1}(h_\Gamma) \right) \text{ a.s.,}\\
(iii) & \sup_{\bm t\in \Omega} |\hat\phi_j (\bm t) -\phi_j (\bm t)|  \\ & = O\left(\zeta + \delta_{n1}(h_\mu)+\delta_{n1}(h_\Gamma) \right) \text{ a.s.,} 
\end{align*}
where $\zeta = h_\mu^2 + h_\Gamma^2+\delta_{n2}^2(h_\Gamma)$, and $J$ is an arbitrary fixed constant.
\end{thm}

Again, we discuss the above results with two different sampling schemes for $d\geq 2$.
\begin{cor} \label{A:cor3}
Assume that A.1-A.6 hold and $1\leq j \leq J$ for an arbitrary fixed constant $J$. \\
(i) Suppose that $\max_{1\leq i\leq n} N_i \leq \mathcal{M}$ for some fixed $\mathcal{M}$. If $ (\log n /n)^{1/2} < h_\mu^d \lesssim h_\Gamma^d \lesssim (\log n /n)^{1/3}$,
\begin{itemize}
\item[(a)] 
$ |\hat\lambda_j -\lambda_j| = O\left(  h_\Gamma^2+ \{\log n/(nh_\Gamma^{2d})\}\right) \text{ } a.s.$;
\item[(b)] both $\sup_{\bm t\in \Omega} |\hat\phi_j (\bm t) -\phi_j (\bm t)|$  and $\|\hat\phi_j(\bm t) -\phi_j (\bm t)\|$ are of the rate $O\left( h_\Gamma^2 + \{\log n/(nh_\Gamma^{2d})\}\right).$
\end{itemize}

\noindent (ii) Suppose $\max_{1\leq i\leq n} N_i \geq \mathcal M_n$, where $\mathcal M_n^{-1} \lesssim h_\Gamma^d,h_\mu^d$ $\lesssim (\log n/n)^{d/4}$ is bounded away from zero. $|\hat\lambda_j -\lambda_j|$, $\sup_{\bm t\in \Omega} |\hat\phi_j (\bm t) -\phi_j (\bm t)|$  and $\|\hat\phi_j(\bm t) -\phi_j (\bm t)\|$ are of the rate $O\left( (\log n/n)^{1/2}\right)$.

\end{cor}

\section{Computational Issues}\label{compissues}

Now, we provide solutions to \textbf{Issues A} and \textbf{B} and introduce GPU (graphics processing unit) parallelization as well as the FFT calculation for local linear smoothers.

\subsection{Matrixization} 

\textbf{Issue A} can be tackled by properly matrixizing $\Gamma(\bm s, \bm t)$ as the following eigen-equations
\begin{equation}
\begin{split}
	&\int_{\Omega}\Gamma(\bm s, \bm t)\phi_{\ell}(\bm s)d\bm s = \lambda_{\ell}\phi_{\ell}(\bm t), \\
	&\int_{\Omega}\phi_{\ell}(\bm t)\phi_\iota( \bm t)d \bm t = 0\mbox{ for all }\ell\neq\iota,
\end{split}
\label{eigendecompose}
\end{equation}
can be well-approximated by their corresponding Riemann sums, which can be represented with matrix operations. We will illustrate the trick with a simple case where $d=2$. Suppose that $\Gamma (\bm s, \bm t)$ is available on a dense regular grid $(s_{1,i},s_{2,j},t_{1,i'},t_{2,j'})$ for $1\leq i,i' \leq M_1$ and $1\leq j,j' \leq M_2$. Let $\bSigma = \left[\Sigma_{ii'}\right]_{1\leq i,i' \leq M_1}$ be a matrix consisting of block matrices $$\Sigma_{ii'}= \big[\Gamma(s_{1,i},s_{2,k},t_{1,i'},t_{2,\ell})\big]_{1\leq k,\ell \leq M_2},$$ and $\bpsi_{\ell}$ be the eigenvector of $\bSigma$ corresponding to eigenvalue $\tilde{\lambda}_{\ell}$. The eigenvalues, $\lambda_\ell$, and eigenfunctions, $\phi_\ell(\bm t)$, can be well-approximated by properly rescaling $\tilde{\lambda}_{\ell}$ and $\bpsi_{\ell}$ when $M_1$ and $M_2$ are sufficiently big. The rescaling is to ensure $\int \phi_\ell^2(\bm t)d\bm t=1$. 
This  matrixization idea holds for general $d$-dimensional cases, and the construction of $\bSigma$ is fairly simple. It is done by reshaping $\Gamma$ to a 2-dimensional matrix; for example, it can be done by the \texttt{reshape} function in MATLAB. Note that if the domain $\Omega$ is not a rectangle, one could simply find a rectangle compact domain $\tilde\Omega \supset \Omega$, and performs integration on it by setting $\Gamma(\bm s, \bm t)=0$ if $\bm s$ or $\bm t$ is not in $\Omega$ (e.g., Chapter 12 in \cite{Stewart:12}).

\subsection{Large-scale local polynomial smoothing}

Computing the local linear estimators, (\ref{eq:muhat})--(\ref{eq:sigma2hatsmooth}), is time-consuming when $N=\sum_{i=1}^{n}N_{i}$ is large. Fortunately, a local linear smoother possesses some key characteristics that make GPU parallelization easy. Now, we take (\ref{eq:muhat}) as an illustrative example. The computations of $\hat\bb$ at any two distinct $\bt$'s are identical, but independent of each other and do not communicate with each other. 
A GPU comprises hundreds to thousands of micro computing units, and each of them can conduct simple calculations simultaneously. However, these computing units do not communicate with each other, and hence GPU parallelization may not be suitable for other smoothing approaches, such as smoothing splines. 
Our numerical experience indicates that the GPU implementation d-FPCA is nearly 100 times faster than the existing MATLAB package on a machine with a cheap GPU (NVIDIA Quadro K600) consisting of 192 threads.

When $N$ is very large, simply considering the GPU paralleled implementation may not be enough since the computational complexities of both (\ref{eq:muhat}) and (\ref{eq:sigma2hatsmooth}) are $O(N^2)$, and that of (\ref{eq:Gammahat}) is $O(N^4)$. Taking the FFT strategy \citep{Wand:94} could greatly reduce the computational complexities of (\ref{eq:muhat})-- (\ref{eq:sigma2hatsmooth}) to $O(N\log N)$, $O(N^2\log N)$ and $O(N\log N)$, respectively. Although the FFT is restricted for regularly observed functional data, the binning strategy \citep{Wand:94} can be taken for irregularly observed data. Even though binned local linear estimators are  asymptotically consistent \citep{HotiH:03}, the consistency highly relies on the order of bin widths. In order to achieve it, the orders of bin widths have to be smaller than those of the optimal bandwidths and \cite{HallW:96} discussed the minimum number of bins so that the binning effect is negligible. To relax the memory limitation of FFT without sacrificing the computational speed, we partition the domain into several overlapping blocks and apply the FFT-based local linear smoothing to each block. We choose overlapping blocks to avoid additional boundary effects due to smoothing. The block size is decided to achieve a balance between system memory and computational speed. 

\subsection{Bandwidth selection}

Bandwidth selection is crucial in smoothing. The bandwidth minimizing the distance between fitted and true functions, i.e. conditional mean integrated squared error (MISE), is often considered as an optimal bandwidth. To avoid overfitting, one commonly considers a cross-validation (CV) procedure to select a proper bandwidth and a valid scheme in FDA is the \emph{leave-one-sample-out} CV due to theoretical considerations. Unfortunately, performing the \emph{leave-one-sample-out} CV is very time-consuming and unlike traditional smoothing problems, the equivalent form in MISE between CV and GCV does not exist. Therefore, a different viewpoint is taken here. The bandwidth selection procedure is to provide some guidance on suitable bandwidths for a given dataset and in practice one may need to further adjust these objective suggestions manually \cite{LiuM:09}. Also, GCV generally generates satisfactory results as the MATLAB package PACE \citep{YaoMW:05a} considers GCV as a default option. As a consequence, we consider the the \emph{leave-one-observation-out} CV to perform bandwidth selection for its nice connection to GCV.

Although the conditional MISE is asymptotically a smooth and convex function of bandwidths, its empirical version could be very noisy and thus with multiple local minima. So, traditional numerical minimization algorithms cannot be directly applied for searching the optimal bandwidths. The CV or GCV procedures in most statistical packages are carried out by grid search; however, grid search is very time-consuming, especially when $d$ is not small. Therefore, we propose a numerical optimization procedure adaptive to the noisy CV or GCV criteria. Our idea originates from the derivative-free trust-region interpolation-based method \citep{MarazziN:02}, a Newton-like algorithm. Here, the gradient and hessian of the objective function are ``estimated'' by locally quadratic interpolations. To be adaptive for the noisy criteria, we replace the interpolation step in \cite{MarazziN:02} with a quadratic regression. Our algorithm requires $O(d^6)$ iterations to converge in the worst case \citep{ConnSV:09}, which is a great improvement over grid search.

\subsection{Random projection}

\textbf{Issue B} could be vital even with a moderate $d$ as the size of $\bSigma$ grows exponentially. Suppose we have $d$-dimensional functional data with 100 observation grids in each coordinate and $\bSigma$ will consist of $10^{4d}$ elements. It is challenging to perform eigendecomposition on $\bSigma$ due to the memory issue. Take $d=3$ for example; it requires approximately $4\times10^{12}\approx4$TB to store $\bSigma$ with single precision floating points. One remedy is employing the random projection approximation \citep{HalkoMT:11}, which is based on the following algebraic property. Suppose $\bSigma=\bPsi\bLambda\bPsi^{T}$ by eigendecomposition. For any $p\times q$ matrix $\bQ$ satisfying $\bQ\bQ^{T}\approx\bI_{p}$ and $q\ll p$, $\bSigma'=\bQ^{T}\bSigma\bQ$ can be expressed as $\bQ^{T}\bPsi\bLambda\bPsi^{T}\bQ$. This implies that one can perform eigendecomposition on a smaller matrix $\bSigma'$ to obtain $\bLambda$ and $\bPsi$. Specifically, $\bPsi\approx\bQ(\bQ^{T}\bPsi)$, where $\bQ^{T}\bPsi$ are the eigenvectors of $\bSigma'$.  To choose a proper $\bQ$, \cite{HalkoMT:11} proposed to employ a random matrix $\tilde{\bQ}=\{\tilde{q}_{ij}\},$ where $\displaystyle\tilde{q}_{ij}\mathop{\sim}^{i.i.d.}N(0,1/q)$, because $E\left(\tilde{\bQ}\tilde{\bQ}^{T}\right)=\bI_{p}$. The computer codes for the random projection approximation are available on the parallel computing platforms, such as Hadoop and Spark. Note that with our block-wise FFT strategy, $\bSigma'$ can be obtained by block matrix multiplications and thus $\bQ^{T}\bPsi$ without storing the entire $\bSigma$ in a computer's memory. However, this random projection strategy is based on the assumption that the rest ($p-q$) eigenvalues are all zeros. If this assumption is violated, we will need to investigate a new algorithm for eigendecomposition and this will be our future project.

Two simulation studies are conducted to demonstrate the performance of our approach. In the first study, we compare the computational speed of our implementations on the local linear smoothing procedure with that of existing statistical packages. Among them, the MATLAB package PACE \citep{YaoMW:05a} is selected as it was specifically designed for one-dimensional FPCA and generates satisfactory results. We consider two different implementations on the local linear smoothing procedures: the GPU parallelization and the FFT strategy. The GPU parallelization implementation is written in C using the OpenACC toolkit \cite{openacc} and called in MATLAB; the FFT implementation is written in MATLAB using the MATLAB subroutine \texttt{fftn}. In the second study, we consider $d=3$ to demonstrate the potential of our approach for 3D functional data. We will demonstrate the computational advantages of taking the random projection strategy. The random projection approximation is implemented in MATLAB on our own.

\subsection{Simulation I}
Following \cite{YaoMW:05a}, we generate the data from 
\begin{equation}
	\begin{split}
		Y_i(t)&=X_i(t)+\epsilon \\
		         &=\mu(t)+\sum_{\ell=1}^{2}A_{i\ell}\phi_{\ell}(t)+\epsilon,
	\end{split}
	\label{sim1d}
\end{equation}
where $\mu(t)=t+\sin(t)$, $\phi_1(t)=-\cos(\pi t/10)/\sqrt{5}$,\\ $\phi_2(t)=\sin(\pi t/10)/\sqrt{5}$, $A_{i1}\sim N(0,4)$, $A_{i2}\sim N(0,1)$, and $\epsilon\sim N(0,0.25)$. The simulation consists of 100 runs and each run contains $100$ random curves sampled from $10000$ equispaced points in $[0,10]$. To have fair comparisons, we select the number of eigenfunctions by setting the FVE to be at least 95\%, and the principal component scores are estimated through integration for both PACE and our d-FPCA. While applying PACE, we choose the data type as ``dense functional'', and all the other options are set as default. To accelerate the computational speed, PACE bins the original data into 400 equispaced bin grids and selects the bandwidths via GCV. 
The experiment is performed on a machine with Intel E3-1230V3 CPU, 32GB RAM and NVIDIA Quadro K600 GPU. The computational time as well as the MISE,
\[
	\frac{1}{100}\sum_{i=1}^{100}\int_{0}^{10}[\hat{X_i}(t)-X_i(t)]^2dt,
\]
are summarized in Table \ref{simtab1}, indicating that our d-FPCA provides slightly (but, not significantly) better reconstructed curves than PACE does, while our computational speed is about 100 times and 300 times faster than PACE with GPU and FFT implementations, respectively. 

\begin{table}[htp]
\caption{\label{simtab1} Computational time (in seconds) and MISE for PACE and d-FPCA (mean$\pm$std).}
\centering
\begin{tabular}{cccc}
\hline
& PACE & d-FPCA (GPU) & d-FPCA (FFT) \\
\hline
Time & 314.1$\pm$11.1 & 3.2$\pm$1.0 & 1.0$\pm$0.3 \\
MISE & .015$\pm$.019 & .014$\pm$.016 & .014$\pm$.016 \\ \hline
\end{tabular}
\end{table}

\subsection{Simulation II}
In the second experiment, we consider the following model
\begin{equation}
	\begin{split}
		Y_i(\bt)&=X_i(\bt)+\epsilon \\
		           &=\mu(\bt)+\sum_{\ell=1}^{4}A_{i\ell}\phi_{\ell}(\bt)+\epsilon,
	\end{split}
	\label{sim3d}
\end{equation}
where \begin{align*}
\mu(\bt) & =\exp\{(\bt-0.5)^T(\bt-0.5)\}, \\
\phi_{\ell}(\bt) & = \prod_{i=1}^3 \{\sin(2\ell\pi t_i)/2\} \text{ for }\ell=1,\dots,4,
\end{align*} 
$\bt=(t_1,t_2,t_3)^T\in[0,1]^3$, $A_{i\ell}\sim N(0,4^{3-\ell})$, and $\epsilon\sim N(0,1/16)$. In each run, we generate $100$ random functions on the nodes of a $64^3$ equispaced grid. Notice that we need $64^6\times4\approx256$ GB RAM to store the entire covariance function $\bSigma$ if single precision floating points are used, and so performing eigendecomposition directly is demanding. Two approaches are compared here.

Both approaches adapt the FFT strategy in the step of estimating $\mu(\bt)$ and $\bSigma$; the difference is that the first approach performs conventional eigendecomposition to obtain eigenfunctions while the other employs the random projection strategy and the blockwise FFT to obtain eigenfunctions. Performing conventional eigendecomposition on $\bSigma$ requires additional $256$ GB of RAM to obtain all the eigenfunctions; clearly, it does not appear feasible in practice. To be conservative in memory usage, we only extract the first 99 eigenfunctions, which is much more than the number of true eigenfunctions. The experiment is carried out on a Microsoft Azure G5 instance, a virtual machine with 32 cores of Intel Xeon E5 V3 CPU and 448GB of RAM. The Microsoft Azure G5 instance is the machine with the largest memories that we can access. In the second approach, the random projection, $\bSigma'=\bQ^{T}\bSigma\bQ$, can be obtained by blockwise matrix multiplications, and hence it does not require $256$ GB RAM to secure $\bSigma'$. To perform blockwise FFT, we partition the covariance function into 8 overlapping blocks. In the random projection step, we use $q=99$. The experiment is carried out on a Microsoft Azure D14 instance, a virtual machine with 16 cores of Intel Xeon E5 V3 CPU and 112GB of RAM.

To save cost, this simulation only consists of 20 runs. To complete one run, the first approach takes approximately 8.5 hours, while the second one takes approximately 5 hours. The integrated squared error (ISE) of the $\ell$th eigenfunction is defined as
\[
	\int_{\Omega}[\hat{\phi}_{\ell}(\bt)-\phi_{\ell}(\bt)]^2d\bt,
\]
and the results are summarized in Table \ref{simtab2}. The reconstruction errors of both methods are almost equivalent, but the one based on the random projection does introduce very little errors into the estimated eigenfunctions. However, those errors are insignificant and negligible compared to its computational advantage. 

\begin{table*}[htp]
\caption{\label{simtab2} The reconstruction errors (mean$\pm$std) and the ISE of estimated eigenfunctions (mean$\pm$std) for d-FPCA with/without random projection (RP).}
\centering
\begin{tabular}{cccccc}
\hline
& Reconstruction & $\hat{\phi}_{1}$ & $\hat{\phi}_{2}$ & $\hat{\phi}_{3}$ & $\hat{\phi}_{4}$ \\
\hline
RP & .003$\pm$6.2$\times10^{-6}$ & .1122$\pm$.0035 & .1137$\pm$.0038 & .1137$\pm$.0038 & .1140$\pm$.0037 \\
w.o. RP & .003$\pm$6$\times10^{-6}$ & .1125$\pm$.0036 & .1133$\pm$.0038 & .1133$\pm$.0038 & .1137$\pm$.0036 \\ \hline
\end{tabular}
\end{table*}

\section{Data Analysis}\label{realdata}

Recent studies have shown PM2.5 is a potential risk factor for lung cancer \citep{RaaschouNielsenetal:13} and heart attack \citep{Giuliaetal:14}; thus, monitoring the level of PM2.5 becomes urgent. 
The dataset consisting of PM2.5 measurements (in $\mu g/m^3$) made at 25 irregularly located monitor sites on western Taiwan from Nov. 29, 2012 to Mar. 1, 2015 (available at \url{http://taqm.epa.gov.tw/pm25/tw/}) is considered in this section. The measurements were carried out manually every three days; therefore, the sample size is 275. Due to experimental failures, machine malfunction and that all the monitor sites were not established simultaneously, the measurements are not complete. One might consider to perform one-dimensional FPCA to this dataset, i.e., the measurements made at each monitor site represent one sample curve. However, doing so leads to serious problems; the sample size becomes very small (i.e., $n=25$) and the samples are highly correlated due to their spatial locations. Moreover, because the measurements were made every three days and they are highly dependent on the weather conditions (e.g., winds and rains), the true temporal correlation is difficult to estimate consistently with current available data resolution. 
All these are problematic in performing FPCA. To the best of our knowledge, no existing functional approach is capable of analyzing such type of data. So, only d-FPCA is applied to this dataset and the following model is considered
\begin{equation*}
Y_i(\bm s) = \mu(\bm s) + \sum_{k=1}^L A_{ik} \phi_k(\bm s) + \epsilon,
\end{equation*}
where $\bm s$ is a given spatial location, $L$ is the number of eigenfunctions and $i=1,\ldots,275$. 

\begin{figure}[htbp]
\subfigtopskip=1pt
\subfigbottomskip=1pt
\centering
\subfigure[]{\includegraphics[width=0.495\linewidth]{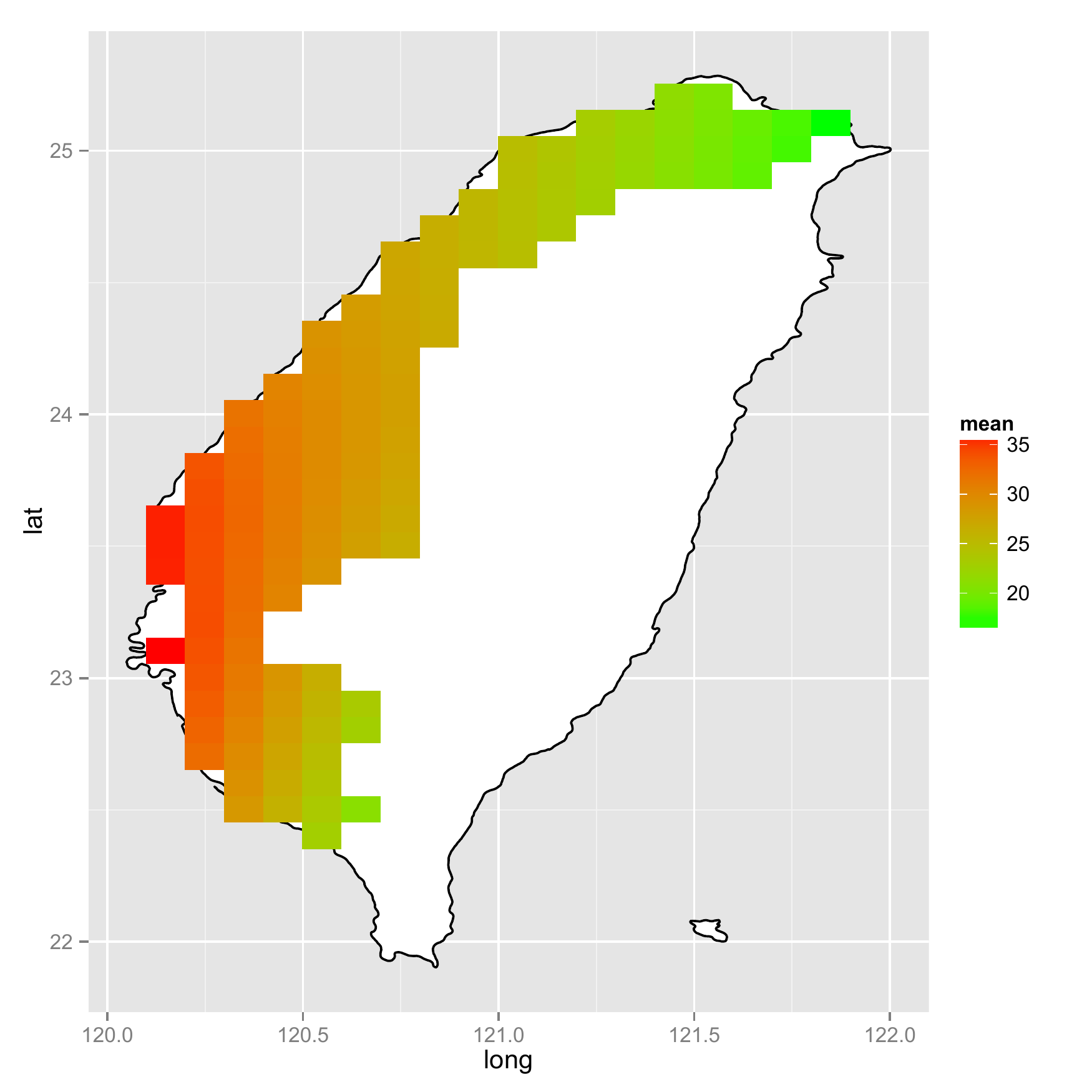}}
\subfigure[]{\includegraphics[width=0.495\linewidth]{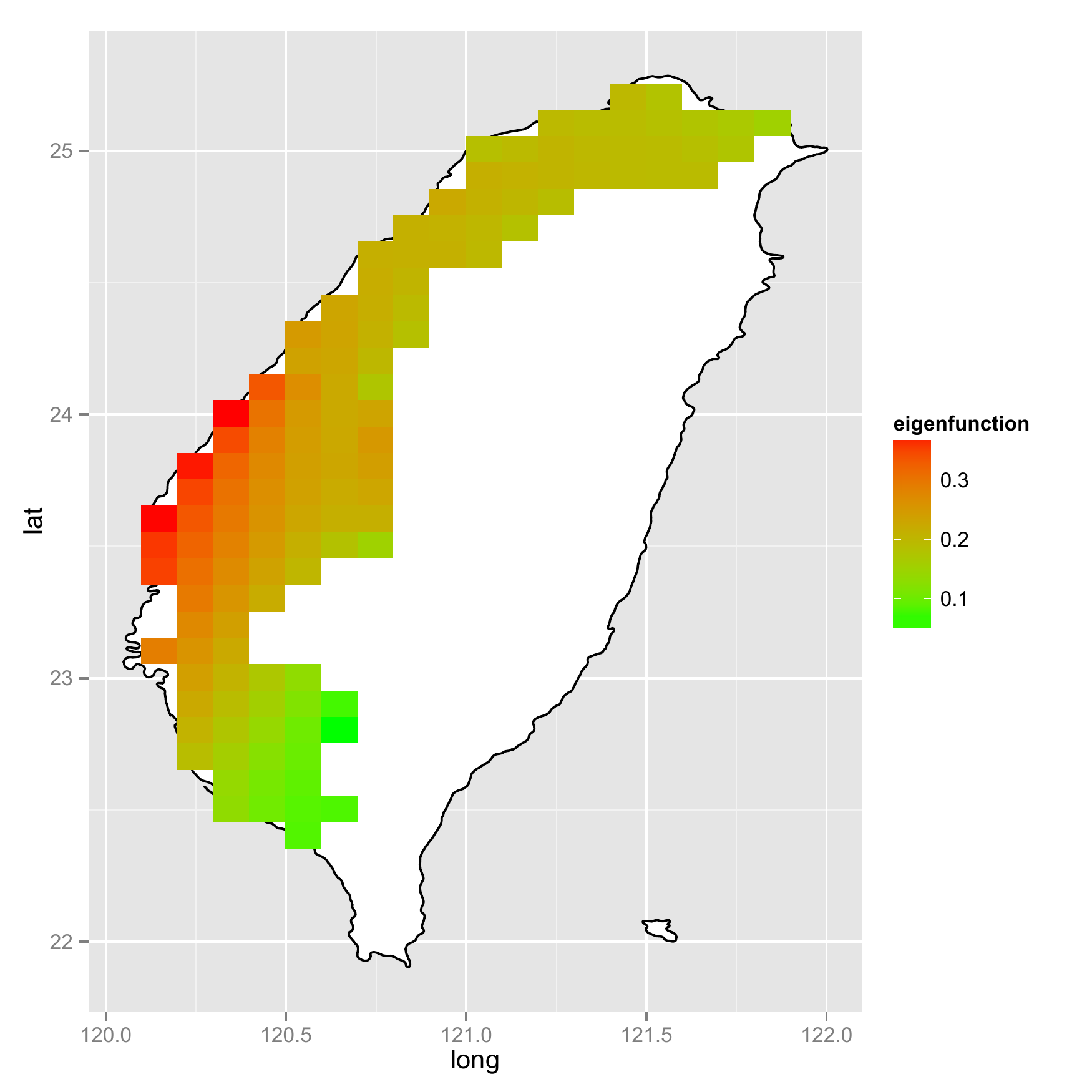}}
\subfigure[]{\includegraphics[width=0.495\linewidth]{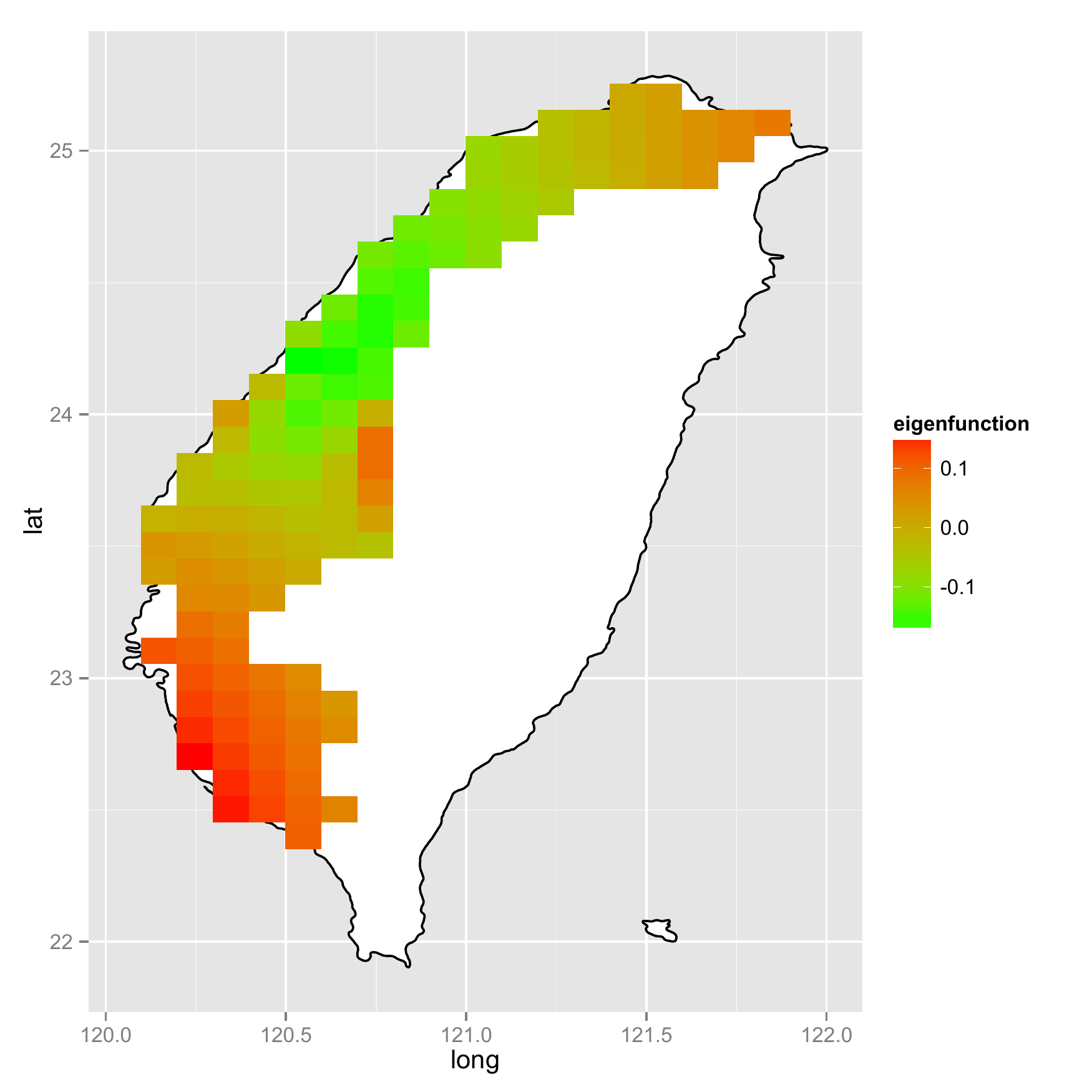}}
\subfigure[]{\includegraphics[width=0.495\linewidth]{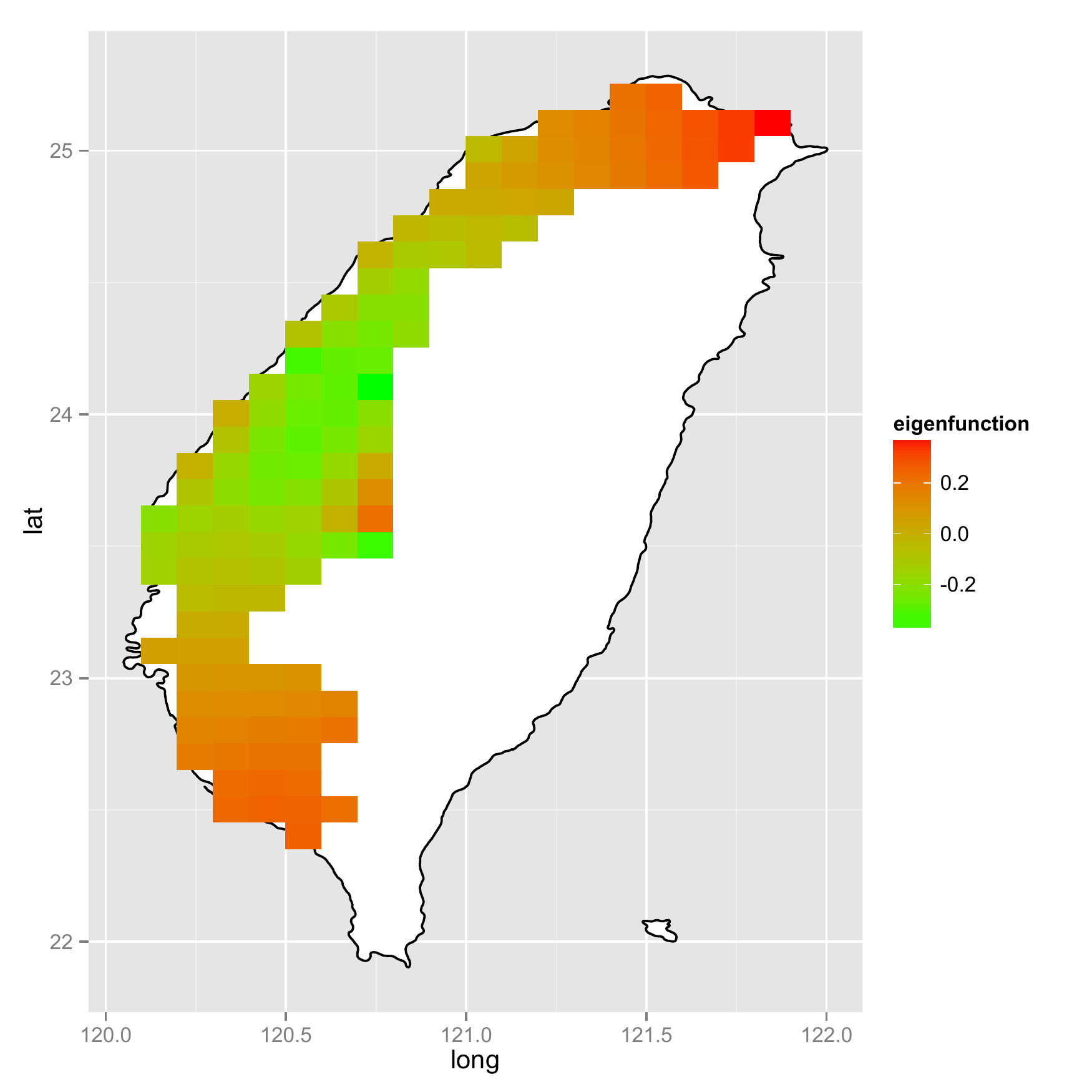}}
\caption{Estimated functions for the PM 2.5 dataset: (a) is the estimated mean function; (b)-(d) correspond to the estimated first three eigenfunctions, respectively.}
\label{pm25eigens}
\end{figure}

\begin{figure}[htbp]
\centering
\includegraphics[width=0.7\linewidth]{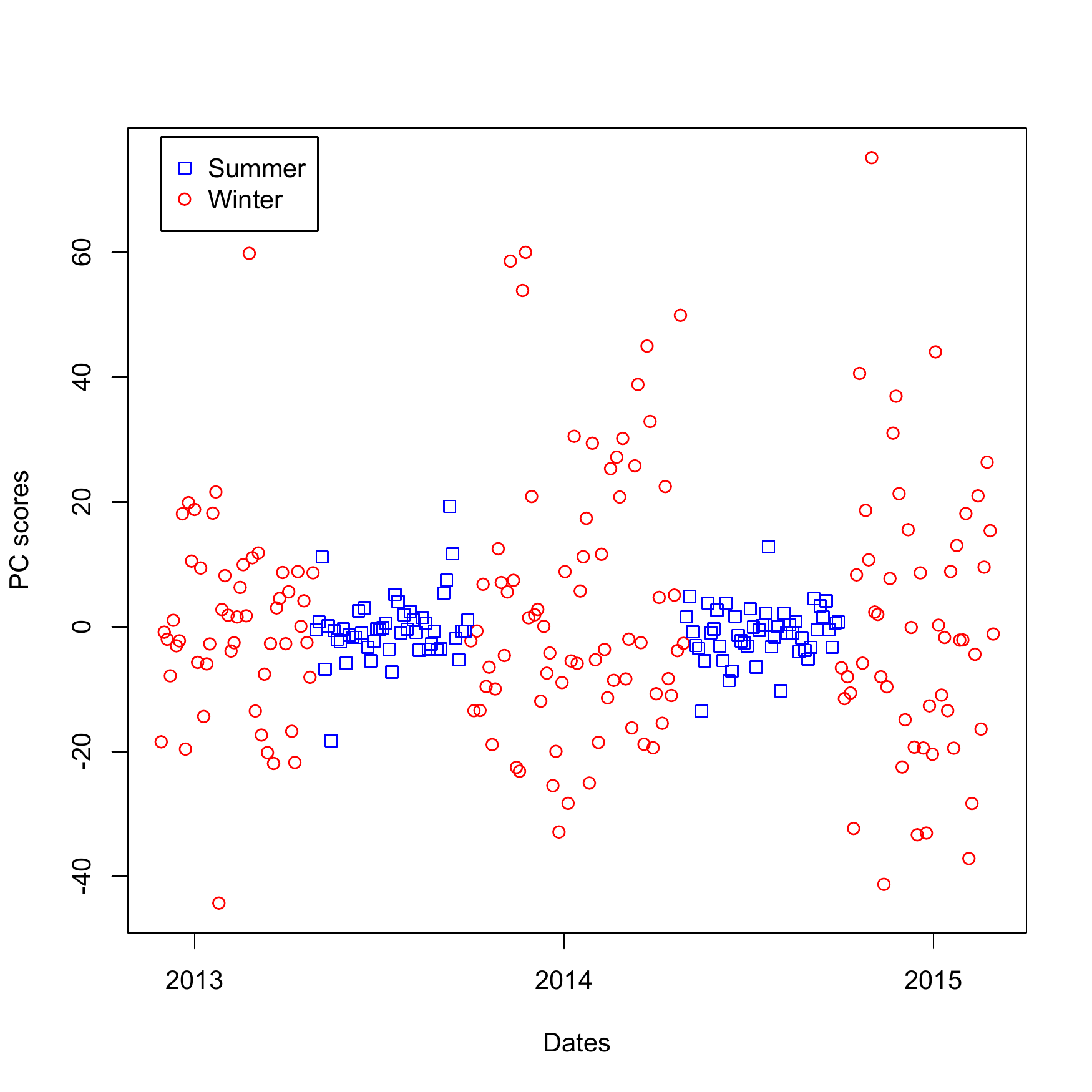}
\caption{The principal component scores of the third eigenfunction.}
\label{pm25_3pcscores}
\end{figure}

Figure \ref{pm25eigens} shows the estimated mean function, $\hat\mu(\bm s)$, and the estimates of the first three eigenfunctions, $\hat\phi_k(\bm s)$ for $k=1,2,3$. The fractions of variation explained by the first three eigenfunctions are 78.6\%, 13.0\%, and 7.1\%, respectively. Figure \ref{pm25eigens}-(a) indicates that the level of PM 2.5 on southwestern Taiwan is higher than that on northern Taiwan on average. This may be because most of the coal thermal power stations and heavy industry factories (e.g., steel mills and petrochemical factories) are on southwestern Taiwan. Note that $|\phi_k(\bm s)|$  represents the variation magnitude of PM 2.5 level at location $\bm s$. A positive $\phi_k(\bm s)$ with a positive $A_{ik}$ implies that the PM 2.5 level at location $\bm s$ at time $i$ is higher than $\mu(\bm s)$ given that other variables remain unchanged. This may be due to the PM 2.5 accumulation at location $\bm s$. On the contrary, a negative $A_{ik}$ implies that PM 2.5 level at time $i$ is lower than its average, which may be because of rains or winds. Figure \ref{pm25eigens}-(b) indicates that the issue of PM 2.5 is very severe around Yunlin County, where the largest petrochemical complex in Taiwan is located, as the first eigenfunction explains 78.6\% variation of the PM 2.5 data and the PM 2.5 level could become very high in this area. Figure \ref{pm25eigens}-(c) indicates that the PM 2.5 levels in Kaohsiung and Tainan (two consecutive cities in southern Taiwan) could become relatively high. One possible explanation is that most of the petrochemical factories in Taiwan are located in Kaohsiung. Figure \ref{pm25eigens}-(d) suggests that the PM 2.5 level in Taipei could also get high even though its average is relatively low and this might result from the air pollutant in China carried by the northeast monsoon in winter. Figure \ref{pm25_3pcscores} further confirms this conjecture as the principal component scores are all very close to zero in summer while they fluctuate a lot in winter. The explanation for negative principal component scores is that it rains very often on northern Taiwan in winter. Further, all three estimated eigenfunctions show that a small area in Nantou county could have relatively high level of PM2.5 and this may be due to the accumulation of fine particulate matter blocked by the Central Range.  

\begin{figure}[htbp]
\subfigtopskip=1pt
\subfigbottomskip=1pt
\centering
\subfigure{\includegraphics[width=0.495\linewidth]{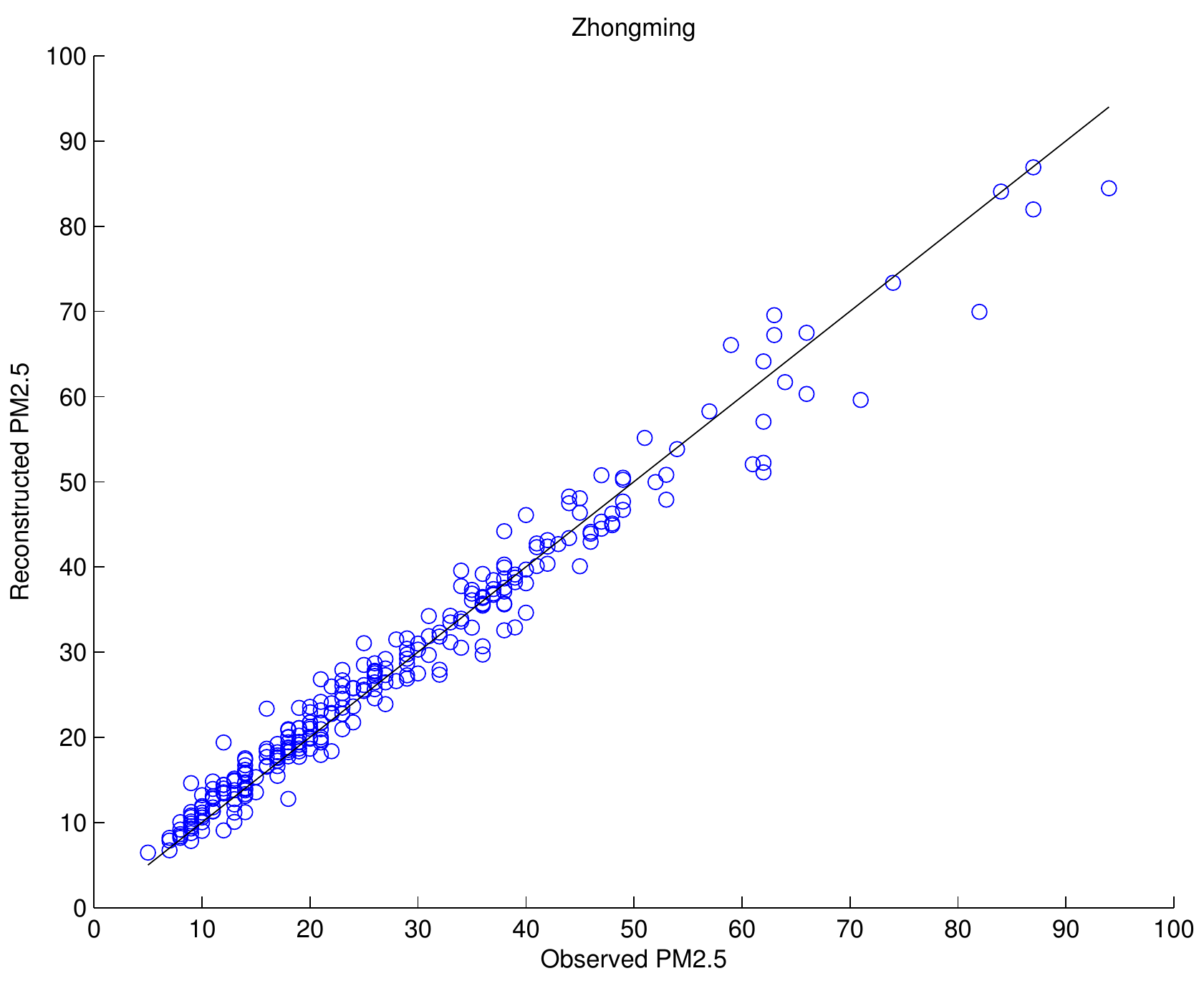}}
\subfigure{\includegraphics[width=0.495\linewidth]{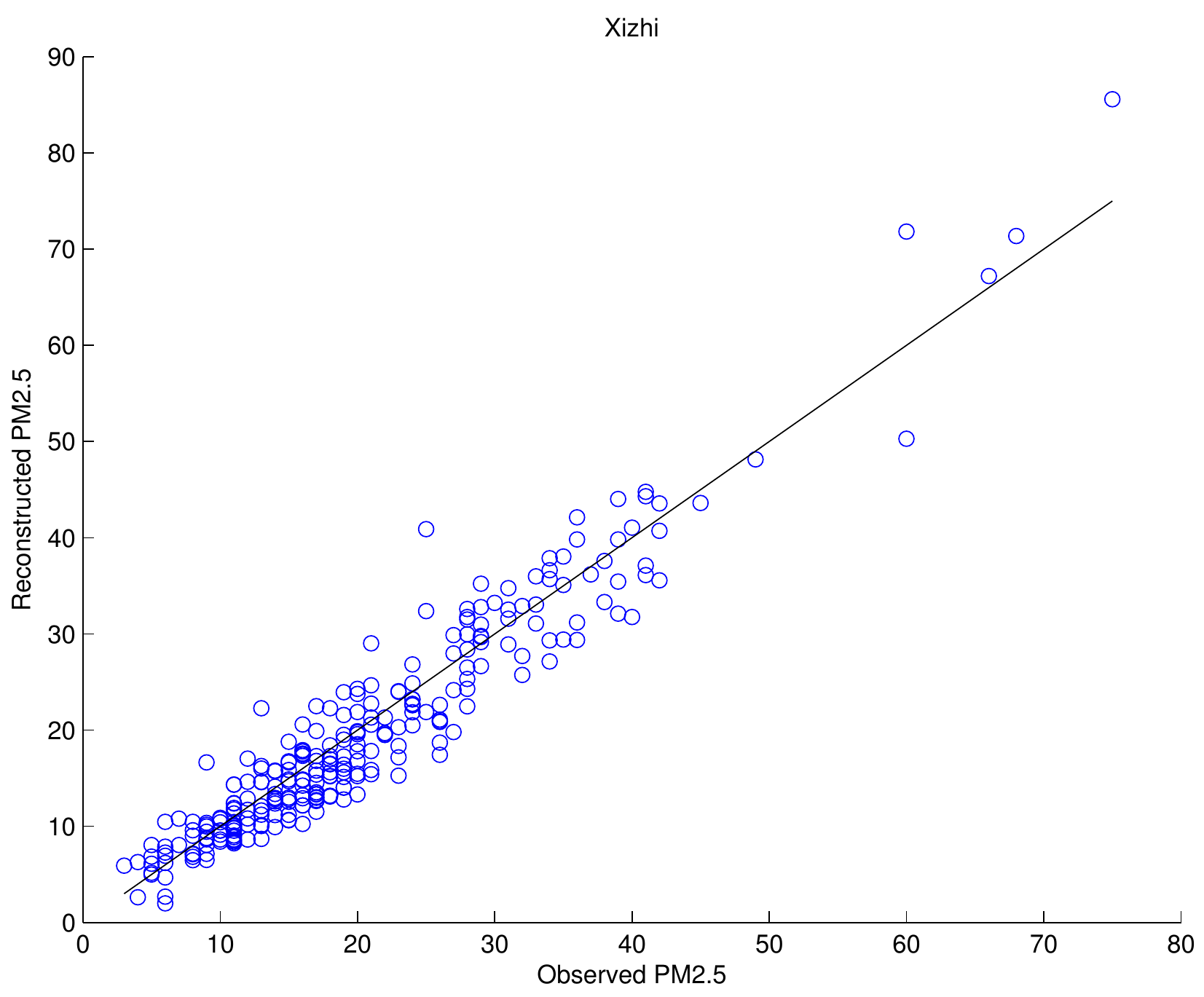}}
\caption{Underlying observations vs Reconstructed PM 2.5 measurements for Zhongming and Xizhi monitor sites.}
\label{pm25}
\end{figure}

Two experiments are conducted to demonstrate the reconstruction ability of d-FPCA. In the first experiment, each run we randomly split the data into training and validation sets with 175 and 100 samples, respectively. It consists of 50 runs. Again, MISE is employed to evaluate the performance of d-FPCA;  
the average and standard error are 29.15 and 4.46, respectively. In the second experiment, we evaluate d-FPCA's ability to predict missing values by means of the \emph{leave-one-location-out} CV strategy. The performance is measured with the squared prediction error, 
\begin{equation}\label{eq:spe2}
	\sum_{j \in \mathcal{I}_i}\left[Y_{j}(\bm s_i)-\hat{X}^{(-i)}_{j}(\bm s_i)\right]^2,
\end{equation}
where $Y_{j}(\bm s_i)$ is the measurement of location $\bm s_i$ made at time $j$, $\hat{X}^{(-i)}_{j}(\bm s_{i})$ is the prediction of $Y_{j}(\bm s_i)$  and $\mathcal{I}_i$ is the set of observation times for location $\bm s_i$. The average and standard error of (\ref{eq:spe2}) are 31.17 and 4.52, respectively. Figure \ref{pm25} shows the reconstructions at two specific locations, the closest stations to where the two authors live. Figure \ref{pm25} indicates that d-FPCA works very well since distributions of the observations and the corresponding predictions are very close to the straight line ($x=y$). 

\section{Conclusions}

The idea of extending FPCA to analyze multidimensional functional data is very natural and the extension is quite straightforwad conceptually and theoretically; however, the emerging computational issues for practical data analyses are not so easy. To tackle these issues, we employed the estimators based on local linear smoothers because of their capabilities of performing large-scale smoothing and handling data with different sampling schemes over various domains in addition to their nice theoretical properties. To accelerate the computational speed, we employed both the FFT strategy and the GPU parallel computing in the smoothing step. The out-of-memory problem in the eigendecomposition step due to large-scale data has been properly addressed by using the random projection approximation and block matrix operations. Our strategies on conducting large-scale smoothing and performing eigendecomposition on huge covariance matrices are completely applicable to other FDA approaches for multi-dimensional functional data, such as partial least-squares \citep{PredaS:05,DelaigleH:12}, functional sliced inverse regression \cite{Ferre03,Ferre05}, etc.

We have also investigated the asymptotic properties of the proposed estimators under two different sampling schemes. Specifically, we have shown that the classical nonparametric rates can be achieved for longitudinal data and the optimal convergence rates can be achieved if $N_i$, the number of observations per sample, is of the order $(n/ \log n)^{d/4}$ for functional data. The finite sample performance of our approach has been demonstrated with simulation studies and the fine particulate matter (PM 2.5) data measured in Taiwan. Although only functional data cases have been considered in our numerical experiments, the proposed approach is definitely applicable for longitudinal data by employing ``PACE'' to predict the principal component scores.

\appendix

\section*{Appendix: Assumptions}

Since the estimators, $\hat\mu(\bm t)$ and $\hat\Gamma(\bm s,\bm t)$, are obtained by applying the local linear smoothing approaches, it is natural to make the standard smoothness assumptions on the second derivatives of $\mu(\bm t)$ and $\Gamma(\bm s, \bm t)$. Assumed that the data
$(\bm T_i,\bm Y_i), i=1,\cdots, n,$ are from the same distribution, where $\bm{T}_i=(\bm t_{i1},\cdots,\bm t_{iN_i})$ and
$\bm{Y}_i=(Y_{i1},\cdots,Y_{iN_i})$.  Notice that $(\bm t_{ij},Y_{ij})$ and $(\bm t_{ik},Y_{ik})$ are dependent but identically distributed and with marginal density $g(\bm t,y)$. Additional assumptions and conditions are listed below.

The following assumptions of $Y(t)$ were also made in \cite{LiH:10:1}.  Suppose the observation of the $i$th subject at time $\bm t_{ij}$ is $Y_{ij} = \mu(\bm t_{ij})+U_{ij}$, where $\text{cov}(U_i(\bm s),U_i(\bm t))=\Gamma(\bm s,\bm t)+\sigma^2 I(s=t)$ and $\Gamma(\bm s,\bm t)=\sum_{\ell} \lambda_\ell \phi_\ell(\bm s)\phi_\ell(\bm t)$.

\begin{itemize}
\item[A.1] For some constant $m_T>0$ and $M_T<\infty$, $m_T\leq g(\bm t,y) \leq M_T$ for all $\bm t\in\Omega$ and $y\in\mathcal{Y}$. Further, $g(\cdot,\cdot)$ is differentiable with a bounded derivative.

\item[A.2] The kernel function $K(\cdot)$ is a symmetric probability density function on $[-1,1]$ and is of bounded variation on $[-1,1]$. 

\item[A.3] $\mu(\bm t)$ is twice differentiable and the second derivative is bounded on $\Omega$.

\item[A.4] $E(|U_{ij}|^{\lambda}) < \infty$ and $E(\sup_{\bm t\in\Omega} |X(\bm t)|^{\lambda}) < \infty$ for some $\lambda \in (2,\infty)$; $h_\mu \rightarrow 0$ and $(h^{2d}_\mu+h^d_\mu/\gamma_{n1})^{-1} (\log n/n)^{1-2/\lambda}$ $\rightarrow 0$ as $n \rightarrow \infty$.

\item[A.5] All second-order partial derivatives of $\Gamma(\bm s, \bm t)$ exist and are bounded on
$\Omega\times\Omega$.

\item[A.6] $E(|U_{ij}|^{2\lambda}) < \infty$ and $E(\sup_{\bm t\in\Omega} |X(t)|^{2\lambda}) < \infty$ for some $\lambda \in (2,\infty)$; $h_\Gamma \rightarrow 0$ and \\ $(h^{4d}_\Gamma+h^{3d}_\Gamma/\gamma_{n1} + h^{2d}_\Gamma/\gamma_{n2})^{-1}(\log n/n)^{1-2/\lambda} \rightarrow 0$ as $n \rightarrow \infty$
\item[A.7] $E(|U_{ij}|^{\lambda}) < \infty$ and $E(\sup_{\bm t\in\Omega} |X(\bm t)|^{\lambda}) < \infty$ for some $\lambda \in (2,\infty)$; $h_\sigma \rightarrow 0$ and $(h^{2d}_\sigma+h^d_\sigma/\gamma_{n1})^{-1} (\log n/n)^{1-2/\lambda}$ $\rightarrow 0$ as $n \rightarrow \infty$.

\end{itemize}

\bibliographystyle{chicago}
\bibliography{FPCA_Image}

\begin{thebibliography}{}

\bibitem[\protect\citeauthoryear{Aston, Chiou, and Evans}{Aston
  et~al.}{2010}]{AstoCE:10}
Aston, J.~A., J.-M. Chiou, and J.~Evans (2010).
\newblock Linguistic pitch analysis using functional principal component mixed
  effect models.
\newblock {\em Journal of the Royal Statistical Society, Series C\/}~{\em 59},
  297--317.

\bibitem[\protect\citeauthoryear{Aston and Kirch}{Aston and
  Kirch}{2012}]{AstoK:12}
Aston, J.~A. and C.~Kirch (2012).
\newblock Estimation of the distribution of change-points with application to
  fmri data.
\newblock {\em Annals of Applied Statistics\/}~{\em 6}, 1906--1948.

\bibitem[\protect\citeauthoryear{Aue, Norinho, and H{\"o}rmann}{Aue
  et~al.}{2015}]{AueDH:14}
Aue, A., D.~D. Norinho, and S.~H{\"o}rmann (2015).
\newblock On the prediction of stationary functional time series.
\newblock {\em Journal of the American Statistical Association\/}~{\em 110},
  378--392.

\bibitem[\protect\citeauthoryear{Breslaw}{Breslaw}{1992}]{Breslaw:92}
Breslaw, J.~A. (1992).
\newblock Kernel estimation with cross-validation using the fast fourier
  transform.
\newblock {\em Economics Letters\/}~{\em 38}, 285--289.

\bibitem[\protect\citeauthoryear{Cesaroni, Forastiere, Stafoggia, Andersen,
  Badaloni, Beelen, Caracciolo, de~Faire, Erbel, Eriksen, Fratiglioni, Galassi,
  Hampel, Heier, Hennig, Hilding, Hoffmann, Houthuijs, J{\"o}ckel, Korek,
  Lanki, Leander, Magnusson, Migliore, Ostenson, Overvad, Pedersen, J, Penell,
  Pershagen, Pyko, Raaschou-Nielsen, Ranzi, Ricceri, Sacerdote, Salomaa, Swart,
  Turunen, Vineis, Weinmayr, Wolf, de~Hoogh, Hoek, Brunekreef, and
  Peters}{Cesaroni et~al.}{2014}]{Giuliaetal:14}
Cesaroni, G., F.~Forastiere, M.~Stafoggia, Z.~J. Andersen, C.~Badaloni,
  R.~Beelen, B.~Caracciolo, U.~de~Faire, R.~Erbel, K.~T. Eriksen,
  L.~Fratiglioni, C.~Galassi, R.~Hampel, M.~Heier, F.~Hennig, A.~Hilding,
  B.~Hoffmann, D.~Houthuijs, K.-H. J{\"o}ckel, M.~Korek, T.~Lanki, K.~Leander,
  P.~K.~E. Magnusson, E.~Migliore, C.-G. Ostenson, K.~Overvad, N.~L. Pedersen,
  J.~P. J, J.~Penell, G.~Pershagen, A.~Pyko, O.~Raaschou-Nielsen, A.~Ranzi,
  F.~Ricceri, C.~Sacerdote, V.~Salomaa, W.~Swart, A.~W. Turunen, P.~Vineis,
  G.~Weinmayr, K.~Wolf, K.~de~Hoogh, G.~Hoek, B.~Brunekreef, and A.~Peters
  (2014).
\newblock Long term exposure to ambient air pollution and incidence of acute
  coronary events: prospective cohort study and meta-analysis in 11 european
  cohorts from the escape project.
\newblock {\em BMJ\/}~{\em 348}.

\bibitem[\protect\citeauthoryear{Chen, Zhang, Petersen, and M\"{u}ller}{Chen
  et~al.}{2015}]{ChenZPM:15}
Chen, K., X.~Zhang, A.~Petersen, and H.-G. M\"{u}ller (2015, Nov).
\newblock Quantifying in nite-dimensional data: Functional data analysis in
  action.
\newblock {\em Statistics in Biosciences\/}~(DOI:10.1007/s12561-015-9137-5),
  1--23.

\bibitem[\protect\citeauthoryear{Chiou}{Chiou}{2012}]{Chiou:12}
Chiou, J.-M. (2012).
\newblock Dynamical functional prediction and classification, with application
  to traffic flow prediction.
\newblock {\em Annals of Applied Statistics\/}~{\em 6}, 1588--1614.

\bibitem[\protect\citeauthoryear{Chiou and Li}{Chiou and Li}{2007}]{ChiouL:07}
Chiou, J.-M. and P.-L. Li (2007).
\newblock Functional clustering and identifying substructures of longitudinal
  data.
\newblock {\em Journal of the Royal Statistical Society, Series B\/}~{\em 69},
  679--699.

\bibitem[\protect\citeauthoryear{Chiou and Li}{Chiou and Li}{2008}]{ChiouL:08}
Chiou, J.-M. and P.-L. Li (2008).
\newblock Correlation-based functional clustering via subspace projection.
\newblock {\em Journal of the American Statistical Association\/}~{\em 103},
  1684--1692.

\bibitem[\protect\citeauthoryear{Chiou and M\"{u}ller}{Chiou and
  M\"{u}ller}{2009}]{ChioM:09}
Chiou, J.-M. and H.-G. M\"{u}ller (2009).
\newblock Modeling hazard rates as functional data for the analysis of cohort
  lifetables and mortality forecasting.
\newblock {\em Journal of American Statistical Association\/}~{\em 104},
  572--585.

\bibitem[\protect\citeauthoryear{Chui and Lai}{Chui and Lai}{1987}]{ChuiL:87}
Chui, C.~K. and M.-J. Lai (1987).
\newblock Computation of box splines and b-splines on triangulations of
  nonuniform rectangular partitions.
\newblock {\em Journal of Approximation Theory and its Application\/}~{\em 3},
  37--62.

\bibitem[\protect\citeauthoryear{Conn, Scheinberg, and Vicente}{Conn
  et~al.}{2009}]{ConnSV:09}
Conn, A.~R., K.~Scheinberg, and L.~N. Vicente (2009).
\newblock Global convergence of general derivative-free trust-region algorithms
  to first- and second-order critical points.
\newblock {\em SIAM Journal on Optimization\/}~{\em 20}, 387--415.

\bibitem[\protect\citeauthoryear{Delaigle and Hall}{Delaigle and
  Hall}{2012}]{DelaigleH:12}
Delaigle, A. and P.~Hall (2012).
\newblock Methodology and theory for partial least squares applied to
  functional data.
\newblock {\em Annals of Statistics\/}~{\em 40}, 322--352.

\bibitem[\protect\citeauthoryear{Fan and Gijbels}{Fan and
  Gijbels}{1996}]{FanG:96}
Fan, J. and I.~Gijbels (1996).
\newblock {\em Local Polynomial Modelling and Its Applications}.
\newblock London: Chapman and Hall.

\bibitem[\protect\citeauthoryear{Ferr\'{e} and Yao}{Ferr\'{e} and
  Yao}{2003}]{Ferre03}
Ferr\'{e}, L. and A.~Yao (2003).
\newblock Functional sliced inverse regression analysis.
\newblock {\em Statistics\/}~{\em 37\/}(6), 475--488.

\bibitem[\protect\citeauthoryear{Ferr\'{e} and Yao}{Ferr\'{e} and
  Yao}{2005}]{Ferre05}
Ferr\'{e}, L. and A.~Yao (2005).
\newblock Smoothed functional inverse regression.
\newblock {\em Statist. Sinica\/}~{\em 15}, 665--683.

\bibitem[\protect\citeauthoryear{Gertheiss, Goldsmith, Crainiceanu, and
  Greven}{Gertheiss et~al.}{2013}]{GertGCG:13}
Gertheiss, J., J.~Goldsmith, C.~Crainiceanu, and S.~Greven (2013).
\newblock Longitudinal scalar-on-functions regression with application to
  tractography data.
\newblock {\em Biostatistics\/}~{\em 14}, 447--461.

\bibitem[\protect\citeauthoryear{Gervini}{Gervini}{2009}]{Gervini:09}
Gervini, D. (2009).
\newblock Detecting and handling outlying trajectories in irregularly sampled
  functional dataset.
\newblock {\em Annals of Applied Statistics\/}~{\em 3}, 1758--1775.

\bibitem[\protect\citeauthoryear{Halko, Martinsson, and Tropp}{Halko
  et~al.}{2011}]{HalkoMT:11}
Halko, N., P.~Martinsson, and J.~Tropp (2011).
\newblock Finding structure with randomness: Probabilistic algorithms for
  constructing approximate matrix decompositions.
\newblock {\em SIAM Review\/}~{\em 53}, 217--288.

\bibitem[\protect\citeauthoryear{Hall and Hosseini-Nasab}{Hall and
  Hosseini-Nasab}{2006}]{HallH:06}
Hall, P. and M.~Hosseini-Nasab (2006).
\newblock On properties of functional principal component analysis.
\newblock {\em Journal of the Royal Statistical Society, Series B\/}~{\em 68},
  109--126.

\bibitem[\protect\citeauthoryear{Hall, M\"{u}ller, and Wang}{Hall
  et~al.}{2006}]{HallMW:06}
Hall, P., H.-G. M\"{u}ller, and J.-L. Wang (2006).
\newblock Properties of principal component methods for functional and
  longitudinal data analysis.
\newblock {\em Annals of Statistics\/}~{\em 34}, 1493--1517.

\bibitem[\protect\citeauthoryear{Hall and Wand}{Hall and Wand}{1996}]{HallW:96}
Hall, P. and M.~P. Wand (1996).
\newblock On the accuracy of binned kernel density estimators.
\newblock {\em Journal of Multivariate Analysis\/}~{\em 56}, 165--184.

\bibitem[\protect\citeauthoryear{Herdman, Gaudin, Perks, Beckingsale,
  Mallinson, and Jarvis}{Herdman et~al.}{2014}]{openacc}
Herdman, J.~A., W.~P. Gaudin, O.~Perks, D.~A. Beckingsale, A.~C. Mallinson, and
  S.~A. Jarvis (2014).
\newblock Achieving portability and performance through openacc.
\newblock In {\em Workshop on Accelerator Programming Using Directives}, pp.\
  19--26.

\bibitem[\protect\citeauthoryear{Hoti and Holmstr\"{o}m}{Hoti and
  Holmstr\"{o}m}{2003}]{HotiH:03}
Hoti, F. and L.~Holmstr\"{o}m (2003).
\newblock On the estimation error in binned local linear regression.
\newblock {\em Journal of Nonparametric Statistics\/}~{\em 15}, 625--642.

\bibitem[\protect\citeauthoryear{Hung, Wu, Tu, and Huang}{Hung
  et~al.}{2012}]{HungWTH12}
Hung, H., P.~Wu, I.~Tu, and S.~Huang (2012).
\newblock On multilinear principal component analysis of order-two tensors.
\newblock {\em Biometrika\/}~{\em 99}, 569--583.

\bibitem[\protect\citeauthoryear{Hyndman and Shang}{Hyndman and
  Shang}{2009}]{HyndmanS:09}
Hyndman, R.~J. and H.~L. Shang (2009).
\newblock Forecasting functional time series.
\newblock {\em Journal of the Korean Statistical Society\/}~{\em 38}, 199--211.

\bibitem[\protect\citeauthoryear{James, Hastie, and Suger}{James
  et~al.}{2000}]{JameHS:00}
James, G.~M., T.~J. Hastie, and C.~A. Suger (2000).
\newblock Principal components models for sparse functional data.
\newblock {\em Biometrika\/}~{\em 87}, 587--602.

\bibitem[\protect\citeauthoryear{Jiang, Aston, and Wang}{Jiang
  et~al.}{2015}]{JianAW:15}
Jiang, C.-R., J.~A. Aston, and J.-L. Wang (2015).
\newblock A functional approach to deconvolve dynamic neuroimaging data.
\newblock {\em Journal of American Statistical Association\/}~{\em (Accepted)}.

\bibitem[\protect\citeauthoryear{Lai and Wang}{Lai and Wang}{2013}]{LaiW:13}
Lai, M.-J. and L.~Wang (2013).
\newblock Bivariate penalized splines for regression.
\newblock {\em Statistica Sinica\/}~{\em 23}, 1399--1417.

\bibitem[\protect\citeauthoryear{Leng and M\"{u}ller}{Leng and
  M\"{u}ller}{2006}]{LengM:06}
Leng, X. and H.~G. M\"{u}ller (2006).
\newblock Classification using functional data analysis for temporal gene
  expression data.
\newblock {\em Bioinformatics\/}~{\em 22}, 68--76.

\bibitem[\protect\citeauthoryear{Li and Hsing}{Li and Hsing}{2010}]{LiH:10:1}
Li, Y. and T.~Hsing (2010).
\newblock Uniform convergence rates for nonparametric regression and principal
  component analysis in functional/longitudinal data.
\newblock {\em Annals of Statistics\/}~{\em 38}, 3321--3351.

\bibitem[\protect\citeauthoryear{Liu and M\"{u}ller}{Liu and
  M\"{u}ller}{2009}]{LiuM:09}
Liu, B. and H.-G. M\"{u}ller (2009).
\newblock Estimating derivatives for samples of sparsely observed functions,
  with application to on-line auction dynamics.
\newblock {\em Journal of American Statistical Association\/}~{\em 104},
  704--717.

\bibitem[\protect\citeauthoryear{Lu, Plataniotis, and Venetsanopoulos}{Lu
  et~al.}{2008}]{LuPV08}
Lu, H., K.~N.~K. Plataniotis, and A.~N. Venetsanopoulos (2008).
\newblock {MPCA}: Multilinear principal component analysis of tensor objects.
\newblock {\em IEEE Transactions on Neural Networks\/}~{\em 19}, 18--39.

\bibitem[\protect\citeauthoryear{Marazzi and Nocedal}{Marazzi and
  Nocedal}{2002}]{MarazziN:02}
Marazzi, M. and J.~Nocedal (2002).
\newblock Wedge trust region methods for derivative free optimization.
\newblock {\em Mathematical Programming\/}~{\em 91}, 289--305.

\bibitem[\protect\citeauthoryear{Preda and Saporta}{Preda and
  Saporta}{2005}]{PredaS:05}
Preda, C. and G.~Saporta (2005).
\newblock {PLS} regression on a stochastic process.
\newblock {\em Computational Statistics and Data Abalysis\/}~{\em 48},
  149--158.

\bibitem[\protect\citeauthoryear{Raaschou-Nielsen, Andersen, Beelen, Samoli,
  Stafoggia, Weinmayr, Hoffmann, Fischer, Nieuwenhuijsen, Brunekreef, Xun,
  Katsouyanni, Dimakopoulou, Sommar, Forsberg, Modig, Oudin, Oftedal, Schwarze,
  Nafstad, Faire, Pedersen, {\"O}stenson, Fratiglioni, Penell, Korek,
  Pershagen, Eriksen, S{\o}rensen, Tj{\o}nneland, Ellermann, Eeftens, Peeters,
  Meliefste, Wang, de~Mesquita, Key, de~Hoogh, Concin, Nagel, Vilier, Grioni,
  Krogh, Tsai, Ricceri, Sacerdote, Galassi, Migliore, Ranzi, Cesaroni,
  Badaloni, Forastiere, Tamayo, Amiano, Dorronsoro, Trichopoulou, Bamia,
  Vineis, and Hoek}{Raaschou-Nielsen et~al.}{2013}]{RaaschouNielsenetal:13}
Raaschou-Nielsen, O., Z.~J. Andersen, R.~Beelen, E.~Samoli, M.~Stafoggia,
  G.~Weinmayr, B.~Hoffmann, P.~Fischer, M.~J. Nieuwenhuijsen, B.~Brunekreef,
  W.~W. Xun, K.~Katsouyanni, K.~Dimakopoulou, J.~Sommar, B.~Forsberg, L.~Modig,
  A.~Oudin, B.~Oftedal, P.~E. Schwarze, P.~Nafstad, U.~D. Faire, N.~L.
  Pedersen, C.-G. {\"O}stenson, L.~Fratiglioni, J.~Penell, M.~Korek,
  G.~Pershagen, K.~T. Eriksen, M.~S{\o}rensen, A.~Tj{\o}nneland, T.~Ellermann,
  M.~Eeftens, P.~H. Peeters, K.~Meliefste, M.~Wang, B.~B. de~Mesquita, T.~J.
  Key, K.~de~Hoogh, H.~Concin, G.~Nagel, A.~Vilier, S.~Grioni, V.~Krogh, M.-Y.
  Tsai, F.~Ricceri, C.~Sacerdote, C.~Galassi, E.~Migliore, A.~Ranzi,
  G.~Cesaroni, C.~Badaloni, F.~Forastiere, I.~Tamayo, P.~Amiano, M.~Dorronsoro,
  A.~Trichopoulou, C.~Bamia, P.~Vineis, and G.~Hoek (2013).
\newblock Air pollution and lung cancer incidence in 17 european cohorts:
  prospective analyses from the european study of cohorts for air pollution
  effects (escape).
\newblock {\em The Lancet Oncology\/}~{\em 14\/}(9), 813 -- 822.

\bibitem[\protect\citeauthoryear{Rice and Wu}{Rice and Wu}{2001}]{RiceW:01}
Rice, J. and C.~Wu (2001).
\newblock Nonparametric mixed effects models for unequally sampled noisy
  curves.
\newblock {\em Biometrics\/}~{\em 57}, 253--259.

\bibitem[\protect\citeauthoryear{Rice and Silverman}{Rice and
  Silverman}{1991}]{RiceS:91}
Rice, J.~A. and B.~W. Silverman (1991).
\newblock Estimating the mean and covariance structure nonparametrically when
  the data are curves.
\newblock {\em Journal of the Royal Statistical Society, Series B\/}~{\em 53},
  233--243.

\bibitem[\protect\citeauthoryear{Risk, Matteson, Ruppert, and andBrian
  S.~Caffo}{Risk et~al.}{2014}]{RiskMREC:14}
Risk, B.~B., D.~S. Matteson, D.~Ruppert, and A.~E. andBrian S.~Caffo (2014).
\newblock An evaluation of independent component analyses with an application
  to resting-state fmri.
\newblock {\em Biometrics\/}~{\em 70}, 224--236.

\bibitem[\protect\citeauthoryear{Ruppert, Sheather, and Wand}{Ruppert
  et~al.}{1995}]{RuppertSW:94}
Ruppert, D., S.~J. Sheather, and M.~Wand (1995).
\newblock An effective bandwidth selector for local least squares regression.
\newblock {\em Journal of American Statistical Association\/}~{\em 90},
  1257--1270.

\bibitem[\protect\citeauthoryear{Silverman}{Silverman}{1982}]{Silverman:82}
Silverman, B.~W. (1982).
\newblock Algorithm {AS} 176: Kernel density estimation using the fast fourier
  transform.
\newblock {\em Journal of the Royal Statistical Society, Series C\/}~{\em 31},
  93--99.

\bibitem[\protect\citeauthoryear{Silverman}{Silverman}{1996}]{Silverman:96}
Silverman, B.~W. (1996).
\newblock Smoothed functional principal components analysis by choice of norm.
\newblock {\em Annals of Statistics\/}~{\em 24}, 1--24.

\bibitem[\protect\citeauthoryear{Stewart}{Stewart}{2012}]{Stewart:12}
Stewart, J. (2012).
\newblock {\em Essential Calculus: Early Transcendentals, 2nd edition}.
\newblock Brooks Cole.

\bibitem[\protect\citeauthoryear{Tian, Huang, Shen, and Li}{Tian
  et~al.}{2012}]{TianHSL:12}
Tian, T.~S., J.~Z. Huang, H.~Shen, and Z.~Li (2012).
\newblock A two-way regularization method for {MEG} source reconstruction.
\newblock {\em Annals of Applied Statistics\/}~{\em 6}, 1021--1046.

\bibitem[\protect\citeauthoryear{Wand}{Wand}{1994}]{Wand:94}
Wand, M.~P. (1994).
\newblock Fast computation of multivariate kernel estimators.
\newblock {\em Journal of Computational and Graphical Statistics\/}~{\em 3},
  433--445.

\bibitem[\protect\citeauthoryear{Wang and Huang}{Wang and
  Huang}{2015}]{WangH15}
Wang, W.-T. and H.-C. Huang (2015).
\newblock Regularized principal component analysis for spatial data.
\newblock {\em arXiv:1501.03221\/}.

\bibitem[\protect\citeauthoryear{Yao, M\"{u}ller, and Wang}{Yao
  et~al.}{2005a}]{YaoMW:05a}
Yao, F., H.-G. M\"{u}ller, and J.-L. Wang (2005a).
\newblock Functional data analysis for sparse longitudinal data.
\newblock {\em Journal of American Statistical Association\/}~{\em 100},
  577--590.

\bibitem[\protect\citeauthoryear{Yao, M\"{u}ller, and Wang}{Yao
  et~al.}{2005b}]{YaoMW:05b}
Yao, F., H.-G. M\"{u}ller, and J.-L. Wang (2005b).
\newblock Functional linear regression analysis for longitudinal data.
\newblock {\em Annals of Statistics\/}~{\em 33}, 2873--2903.

\bibitem[\protect\citeauthoryear{Zhang, Li, Beckes, and Coan}{Zhang
  et~al.}{2013}]{ZhanLBC:13}
Zhang, T., F.~Li, L.~Beckes, and J.~A. Coan (2013).
\newblock A semi-parametric model of the hemodynamic response for multi-subject
  fmri data.
\newblock {\em NeuroImage\/}~{\em 75}, 136--145.

\bibitem[\protect\citeauthoryear{Zhou and Pan}{Zhou and Pan}{2014}]{Zhoup14}
Zhou, L. and H.~Pan (2014).
\newblock Principal component analysis of two-dimensional functional data.
\newblock {\em Journal of Computational and Graphical Statistics\/}~{\em 23},
  779--801.

\bibitem[\protect\citeauthoryear{Zhu, Zhang, Ibrahim, and Peterson}{Zhu
  et~al.}{2007}]{ZhuZIP:07}
Zhu, H., H.~Zhang, J.~Ibrahim, and B.~Peterson (2007).
\newblock Statistical analysis of diffusion tensors in diffusion-weighted
  magnetic resonance image data (with discussion).
\newblock {\em Journal of American Statistical Association\/}~{\em 102},
  1081--1110.

\bibitem[\protect\citeauthoryear{Zipunnikov, Caffo, Yousem, Davatzikos,
  Schwartz, and Crainiceanu}{Zipunnikov et~al.}{2011a}]{ZipunnikovCYDSC11a}
Zipunnikov, V., B.~Caffo, D.~M. Yousem, C.~Davatzikos, B.~S. Schwartz, and
  C.~Crainiceanu (2011a).
\newblock Functional principal component model for high-dimensional brain
  imaging.
\newblock {\em NeuroImage\/}~{\em 58}, 772--784.

\bibitem[\protect\citeauthoryear{Zipunnikov, Caffo, Yousem, Davatzikos,
  Schwartz, and Crainiceanu}{Zipunnikov et~al.}{2011b}]{ZipunnikovCYDSC11b}
Zipunnikov, V., B.~Caffo, D.~M. Yousem, C.~Davatzikos, B.~S. Schwartz, and
  C.~Crainiceanu (2011b).
\newblock Multilevel functional principal component analysis for
  high-dimensional data.
\newblock {\em Journal of Computational and Graphical Statistics\/}~{\em 20},
  852--873.

\end{thebibliography}

\label{LastPage}

\end{document}